\documentclass{kluwer}    
\input psfig.sty

\begin{document}                                                                                   
\begin{article}
\begin{opening}         
\title{Interpretation of Extreme Scattering Events} 
\author{Mark A. Walker}  
\runningauthor{M. A. Walker}
\runningtitle{Interpretation of ESEs}
\institute{Research Centre for Theoretical Astrophysics,
School of Physics A28,\\
University of Sydney, NSW 2006, Australia}
\date{July 17th, 2000}

\begin{abstract}
Extreme Scattering Events are sometimes manifest in the
light-curves of compact radio-quasars at frequencies of
a few GHz. These events are not understood. The model
which appears to offer the best explanation requires a
new population of AU-sized, neutral gas clouds; these clouds
would then make up a large fraction of the Galaxy's dark matter.
Independent of the question of which theoretical model is
correct, if we extrapolate the observed behaviour to low
radio-frequencies, we expect that the sky should be criss-crossed
by a network of narrow caustics, at frequencies below
about 700~MHz. Consequently at these frequencies sources should
typically manifest additional, faint images which are substantially
delayed with respect to the primary image.
Although some examples of this type of
behaviour are already known, it is expected that these are
just the tip of the iceberg, with strong selection biases
having been imposed by the instrumentation employed to date.
 
\end{abstract}

\end{opening}           

\section{Introduction}  
Extreme Scattering Events (ESEs), Fiedler et al (1987: F87), were
discovered more than a decade ago but are still not understood.
It is generally agreed that these events are due to refraction
by intervening, ionised, Galactic gas (e.g. F87; Romani, Blandford
\& Cordes 1987: RBC87), but there is no agreement on
the astrophysical context in which this gas arises. Most models
attempt to explain ESEs with the minimum possible extrapolation
from conventional astrophysical pictures (see, especially,
Deshpande \& Radhakrishnan 2000); this is the most conservative
approach, and it seems very likely that at least one of these
conventional models will prove relevant. 
There are, however, real difficulties in trying to explain
some of the observed events -- in particular the ESE in Q0954+658
-- with conventional astrophysics, and this has motivated one
rather exotic model in which a new population of dense,
neutral gas clouds is invoked (Walker \& Wardle 1998).

The current lack of consensus on the correct physical picture
for ESEs persists principally because the existing data have
a fairly low information content.  At this point the field is
badly in need of some new observational initiatives; some ideas
are presented in \S6 (see also Walker 2000).

\section{Basic constraints on lenses}
The refracting structures which give rise to ESEs are conveniently
referred to as ``lenses'', although we should bear in mind that they
might not be well-defined physical entities (model (iii) of \S5). There
are three basic properties of the individual lenses which are dictated
fairly directly by the data on ESEs: (i) their transverse dimensions
should be a few AU, (ii) the peak electron column-density
should be of order $10^{17}\;{\rm cm^{-2}}$, and (iii) they should
be symmetric. Point (i) follows immediately from the observed event
durations (months) together with an assumed transverse speed of
order $10^2\;{\rm km\,s^{-1}}$. Point (ii) is deduced by requiring
a strong lens which can magnify a large fraction of a source which
is of order a milli-arcsecond in size. These points were recognised
at the time of discovery of the ESEs (F87). One detail
deserves clarification however: an upper limit on the distance of
the lenses follows from their transverse dimensions in combination
with the requirement that they be larger in angular size than the
source. This reasoning is correct, but the angular size of the
source has previously been taken as the scatter-broadened size, leading
to a distance upper limit of order one kpc, and this is overly
restrictive. For distances of a kpc or more, at high Galactic
latitude, the lens is beyond the majority of the scattering material
in the Galactic disk, and the relevant angular size is then the
intrinsic source size; this can be substantially smaller than the
scatter-broadened size, thereby relaxing the distance limit.

The third point has not previously been emphasised; it
arises simply because the ESE light-curves are, crudely
speaking, time-symmetric. At first sight this statement
appears to have little value, because of the qualifying
phrase ``crudely speaking'', but this is not the case
--- most of the models which have been proposed for ESEs
incorporate no lens symmetry whatsoever, and are therefore
not good starting points for explaining even an approximate
time-symmetry. That's not to say that such models are
excluded, because it might be possible to construct versions
in which the lenses do yield such behaviour, but the point remains
that this property must be explained somehow. Some symmetry
might be effected by the process of averaging over the source
structure, but this is true only for the angular/temporal scales
corresponding to the source size, below which flux variations
are suppressed. One might argue that
models which involve symmetric lenses should, in turn, explain
why the observed time-symmetry is only approximate. This, of course,
is trivial, because real astrophysical entities never conform
exactly to the symmetries which are employed in modelling them.

Finally, if the approximate time-symmetry of the ESE light-curves
is not accidental, it requires that any
straight line drawn across the lens plane (representing the apparent
path of the background source) should manifest a reflection symmetry
about one point. In turn this indicates that the lens itself should
have either mirror-symmetry and translational-invariance, or else
it should be axisymmetric.

\section{Further constraints: the case of Q0954+658}
The ESE observed in Q0954+658 (F87) is by far
the most spectacular event observed to date and deserves
particular attention. This event exhibits a number of sharp
peaks in the high-frequency (8.1~GHz) light-curve; these
peaks are generally interpreted as being due to caustics.
While there are four large peaks
evident, there are roughly seven smaller, sharp peaks in this
same light-curve, making eleven in total. Now caustic curves
are closed curves, so that during a lensing event a source
which crosses from the exterior to the interior of this
boundary must later cross to the exterior again, giving
rise to two peaks in the light curve. Furthermore, for a
diverging lens the caustic curves come in pairs --- one
pair for every peak in electron column-density (provided
the peak is sufficiently sharp). Thus, even if the source
structure is a single-component only,
the 8.1~GHz light-curve could be reproduced with as few as
three column-density peaks, implying that the column
density profile of the lens is likely to be very simple.

\section{Nature of the lens symmetry}
It is straightforward to decide which of the two possible
lens symmetries (\S2) is preferred; it is the axisymmetric
lens. This can be seen immediately from figure 1, which
shows examples of the low-frequency light-curves arising
from axisymmetric/mirror-symmetric lenses for which, in
both cases, the source passes behind two peaks in electron
column-density. (The mirror-symmetric lens consists of
two parallel filaments, with Gaussian cross-sections, while
the axisymmetric lens is a simple ring, again with a Gaussian
cross-section.) A single Gaussian component is used for
the source structure in these calculations. Caustic
crossings are seen as the peaks in the
light-curves; only seven are visible because the
central peak contains an unresolved pair in each case.
Both light-curves exhibit the same deep flux
depression, when the source is nearly on-axis. Conservation
of energy demands that this power appears somewhere
else in the observer's plane, and it is in this respect
that the two lenses differ greatly: for the
translationally-invariant lens the light-curve actually 
manifests this flux conservation, in the sense that
the flux averaged over the whole event is equal to the
unlensed flux, whereas this is not true for the axisymmetric
lens. The data for Q0954+658 clearly favour the axisymmetric
model.\goodbreak
\centerline{\psfig{figure=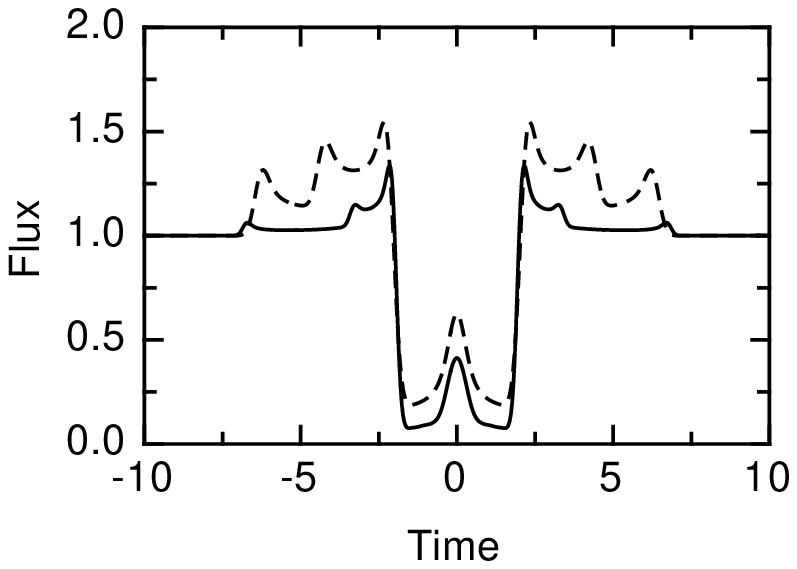,width=8cm}}
\centerline{{\bf Figure 1.\/} Low frequency light-curves for ESEs produced by}
\centerline{mirror-symmetric (dashed) and axisymmetric (solid) lenses.}

\section{Overview of models}
A number of models have been proposed to explain ESEs,
in which the lenses are identified with a variety of
physical phenomena. It should be borne in mind that
the defining criteria for ESEs (F87, Fiedler
et al 1994: F94) have been very loosely framed (``periods
of unusual variability''), and this may have created
heterogeneity in the class --- more than one phenomenon could
be represented amongst the events which have been dubbed ESEs.
More than one lens model may therefore be relevant.
A brief summary of lens models follows:\\
(i) Random refracting elements (F87, F94)\\
(ii) Magnetically confined filaments (RBC87)\\
(iii) Steep spectrum turbulence (Deshpande 
 \& Radhakrishnan 2000)\\
(iv) Shock waves (RBC87; Clegg, Chernoff 
 \& Cordes 1988)\\
(v) Photo-ionised surfaces of giant clouds (Rickett, Lyne \& Gupta 1997)\\
(vi) Photo-ionised winds from AU-sized clouds (Walker \& Wardle 1998).\\
Of these models, (i) is at present a purely phenomenological model
whose physical viability cannot be readily assessed. Models (ii)
and (iii) possess no particular symmetry, and are therefore disfavoured,
while models (iv) and (v) involve strongly asymmetric lenses
and are strongly disfavoured in this respect; only model (vi)
generates the observed quasi-symmetry in a natural way. The
necessary electron column densities and scale-sizes
may in principle be realised by any of models (ii--vi), but models
(ii), (iv) and (v) need to be developed further before meaningful
assessments can be made. Model (vi) appeared, initially, to yield
the necessary column/scale-size combination in a very natural way,
but McKee (2000) has since pointed out
that a photo-evaporated wind would, in this context, have a modest
ionisation fraction, so the calculation of the ionised column-density
needs to be revisited for this model.

Notwithstanding a huge reduction in the predicted column
of ionised gas, model (vi) currently appears to offer the best
explanation for some of the ESEs -- notably Q0954+658
-- and the main issue is whether or not a population of dense,
neutral clouds actually exists. Indeed this is a question with
ramifications throughout astrophysics, because the neutral
clouds would have to constitute a major component of the
Galactic dark matter. The putative clouds cannot
be excluded on the basis of any existing data (Walker \& Wardle
1999), and this model provides a strong motivation
for intensive study of the ESE phenomenon.

\section{Future work}
How can we make progress in this field? A key aspect of the problem is
the fact that ESEs are rare.
This difficulty can most easily be addressed by working at low
frequencies, where the refraction angles are larger and the
cross-section for multiple imaging is increased. Indeed,
for a lens which is localised in both transverse dimensions,
the optical depth for multiple-imaging should scale as
$\lambda^4$, independent of the actual lens model. Taking
the optical depth for Extreme Scattering (extragalactic
sources) to be of order $5\times10^{-3}$ at 2.7~GHz (F94),
it is straightforward to predict that at frequencies
below about 700~MHz there will be multiple images present
most of the time. At these frequencies, then, the sky should
exhibit a network of caustics. This does {\it not\/} mean
that large flux changes will be happening continuously
below this frequency, because the caustics are very narrow and
in total cover only a tiny fraction of the sky; rather it means
that there should typically
be some extra {\it faint\/} images present. This phenomenon
is, in fact, well known from pulsar studies (Cordes \& Wolszczan
1986; Rickett 1990), where it manifests itself as interference
fringes in the dynamic spectra; it is also very common,
occurring for more than 10\% of the time for some pulsars
(J.M.~Cordes, 2000, personal communication). However, while
the connection to ESEs has long been recognised (e.g. RBC87),
the exact relationship between
the two effects remains to be understood. Regrettably, the
multiple imaging phenomenon has not yet been exploited in
any systematic way to learn about the lenses.

There are a number of possible avenues to improving the current
situation by working with these multiple images. For example:
one could gain some information on the structure of the lens
simply by counting the number of images present; lens symmetry
could be studied via VLBI observations through the
course of a multiple imaging event; for long-duration events
the evolution of the image delays could yield a ``parallax''
measurement; and magnetic fields in the lenses
could be studied by comparing the fringe patterns in different
polarisations.

It is important to note that the faint, ``extra'' images can
be substantially delayed with respect to the main image,
and the magnitude of the delay is roughly proportional to
the geometric area covered by the images, hence proportional
to the optical depth. Now refraction through an angle of
order a milli-arcsecond should, over a distance of order
one kpc, introduce a geometric delay of order $10^{-6}$~sec.
Thus for lenses of order a milli-arcsecond in size, a strong
lensing event (ESE) should introduce image delays of this
magnitude, while multiple imaging at frequencies below
700~MHz will introduce delays hundreds of times larger.
Such images would be extremely difficult to
detect with conventional techniques, because the interference
fringes would be so fine that they could not be resolved
with existing spectrometers. It is therefore to be expected
that observations of mutiple imaging phenomena have,
to date, been subject to strong instrumental biases which
allow us to see only images with relatively small delays,
with the typical secondary images being censored. This
bias increases in severity very rapidly as the observing
frequency is decreased, with the maximum delay scaling roughly
as $\lambda^4$ in this regime. To be confident that we
are not introducing a bias, the
only way forward appears to be the use of base-band
recording, from which the temporal auto-correlation of the
electric field, for example, can be computed out to large lags.

\end{article}
\end{document}